\documentstyle[12pt,mont98_desy,amssymb,epsfig]{article}

\begin{document}

\title{\large{\bf{Spin transport, spin diffusion and Bloch equations in
electron storage rings}}
\footnote{Extended version of a contribution to the proceedings of the workshop
``Nonlinear problems in charged beam transport in linear and recirculating
accelerators'', ENEA, Frascati, Italy, May 1998 and of
DESY Report 98--145.}
}
\author{K. Heinemann and {{D.P. Barber}}}

\address{Deutsches Elektronen--Synchrotron, DESY, \\
 22603 Hamburg, Germany. \\E-mail: heineman@mail.desy.de,
mpybar@mail.desy.de}

\maketitle

\begin{abstract}
We show how, beginning with the Fokker--Planck equation for electrons 
emitting synchrotron radiation in a storage ring, the corresponding equation  
for spin motion can be constructed. This is an equation of the Bloch type
for the polarisation density.  
\end{abstract}

\section{\bf Introduction and motivation.}
Relativistic electrons circulating in a
storage ring emit synchrotron radiation and a tiny fraction of the
photons can cause spin flip from up to down and vice versa. However, the
$ \uparrow\downarrow $ and $\downarrow\uparrow$  rates differ so that 
the beam can become spin polarised antiparallel to the guide field, reaching a 
maximum polarisation, $P_{_{\rm ST}}$, of ${8}/(5\sqrt{3}) = 0.924$ 
in a perfectly aligned flat ring without solenoids. This is the so--called
Sokolov--Ternov (ST) effect \cite{st64}.
The time constant  for the exponential  build--up is
usually in the range of a few minutes to a few hours and decreases with the
fifth power of the energy.

However real rings can have misalignments, solenoids and spin rotators and 
this together with the fact that spins precess in magnetic fields
has important consequences for the polarisation which we now outline.

In the absence of spin flip, spin motion for electrons moving 
in electric and magnetic fields is
described by the T--BMT   equation \cite{th27,bmt59}
{}~~$d \vec {S}/{ds}= \vec {\Omega} \wedge \vec {S} $  
where $\vec S$ is the rest frame spin expectation value of the
electron, ~$s$ is the distance around the ring
         and $\vec{\Omega}$ depends on the electric and magnetic fields,
the velocity and the energy. In  magnetic fields 
{}~~$\vec {\Omega} = -e/ ({mc \gamma})
\left[( 1 + a) \vec {B}_{\|} + (1 + a \gamma) \vec {B}_{\perp}\right]$ 
where $\vec {B}_{\|}$ and $\vec {B}_{\perp}$ are the laboratory magnetic
fields parallel and perpendicular to the trajectory.
The gyromagnetic anomaly, $a = (g - 2)/{2}$, for electrons is
about 0.001159652. For protons it is about 1.7928. 
The other symbols have their usual meanings.
Thus for motion transverse to the
magnetic field, the spin precesses around the field at a rate $(1 +a \gamma)$ 
faster than the rate of change of the orbit direction \cite{chao81,
chao82}.
For electrons at 27.5 GeV at HERA \cite{ba95} the 
{\it spin enhancement factor } $a \gamma$, is about 62.5.  

Synchrotron radiation not only creates polarisation
but also produces other effects. In particular, the stochastic element of
photon emission together with accompanying damping determines the equilibrium
phase space density distribution and the beam  can be described by a 
Fokker--Planck (FP) equation. This is traditionally derived by simulating
the stochastic photon emission with Gaussian white noise 
\cite{bhmr91,jow87,rug90,risken}.
The same photon emission also imparts a
stochastic element to $\vec{\Omega}$  and then, via the T-BMT
equation, spin diffusion (and thus depolarisation) can occur in the
inhomogeneous fields of the ring \cite{baiorl66,mont98I,br98}.
Thus synchrotron radiation can create polarisation but can also  lead to its
destruction! 
The ratio (depolarisation rate / polarisation rate) increases with the spin 
enhancement factor.
The equilibrium  
polarisation is the result of a balance between the Sokolov-Ternov effect
and this radiative depolarisation so that the attainable polarisation
$P_{\rm eq}$ is less than $P_{_{\rm ST}}$.  In the approximation that
the orbital motion is linear, the {\it value} of the polarisation is
essentially the same at each point in phase space and in $s$ and the 
polarisation is aligned along the Derbenev-Kondratenko vector 
$\hat n$ \cite{dk73}.
The unit vector {\it field}  $\hat{n}$
 depends on $s$ and the position in phase space 
defined by  $\vec u \equiv(x, p_x, y, p_y, \Delta s, \delta =  \Delta E /E_0)$
\cite{bhr1}.
${\hat n}({\vec u(s)};s)$ satisfies the T-BMT equation along any orbit
$\vec u(s)$ and it is periodic in azimuth: 
$\hat{n}(\vec{u}; s)   =  \hat{n}(\vec{u}; s+C)$ where $C$ is the ring
circumference.
On the closed orbit ($\vec u = \vec 0$), $\hat{n}(\vec{u}; s)$
is denoted by $\hat{n}_{0}(s)$.

Taking into account radiative depolarisation
due to photon-induced longitudinal recoils, the equilibrium electron 
polarisation along the $\hat{n}$ field as given by 
Derbenev and Kondratenko and by Mane is \cite{dk73,mane87}
\begin{eqnarray}
     {P}_{\rm dkm} =
        -\frac{8}{5\sqrt{3}}
     \frac{
   {\oint {ds}            \left<  |K(s)|^{3}
              \hat{b}
              \cdot
      (
       \hat{n}-
                 \frac{\partial{\hat{n}}}
                      {\partial{\delta}}
                           )
          \right>_{s}
                                    }
                                      }
          {
   { \oint {ds}           \left<  |K(s)|^{3}
          ( 1-
              \frac{2}{9}
     { (
       \hat{n}\cdot\hat{s}
                           )}^{2}
              +
      \frac{11}{18}
      \left(
                 \frac{\partial{\hat{n}}}
                      {\partial{\delta}}
                                        \right)^{2} \, )
          \right>_{s}
                                    }
                                     }  
\label{eq:PDK}
\end{eqnarray}
where $\hat{b}$ and $\hat s$ denote the magnetic field direction and direction
of motion respectively, and where $<\ >_{s}$ denotes an average over phase
space at azimuth $s$. The quantity $K(s)$ is the orbit curvature due to the
magnetic fields.
The term $\frac{11}{18}(\partial{\hat{n}}/{\partial{\delta}})^2$,  encapsulates
the spin diffusion.  The ensemble average of the polarisation is
$  { \vec  P}_{\rm eq}(s)  =  P_{\rm dkm}~\langle \hat{n} \rangle_{s}$
and $ \langle \hat{n}\rangle_{s}$ is very nearly aligned along
${{\hat n}_0}(s)$. For the perfectly aligned flat ring without solenoids
mentioned at the beginning,  $ \partial \hat{n}/\partial \delta$ vanishes
so that $P_{\rm dkm} = 0.924$. Further details on this formalism can be found 
in
\cite{mont98I,br98}. 

The Derbenev--Kondratenko--Mane (DKM) formula is based on the reasonable
and justifiable assumption that at spin--orbit equilibrium the polarisation
is locally essentially parallel to $\hat n$ \cite{mont98I}.
But it would be
more satisfying to have access to a more basic approach  free of
assumptions. For example, it would be good to have a kind of spin--orbit
FP equation which would allow non--equilibrium spin--orbit systems
to be studied, and to be able to obtain the DKM result as a special 
case. An
equation of this type for spin  has already been derived by Derbenev and
Kondratenko \cite{dk75}  using semiclassical radiation theory beginning
with the density operator for the spin--orbit system.
This equation includes the effects of spin 
diffusion, the ST effect and also some ``cross terms''.
At the same time they obtained the FP equation for the orbital 
motion  with the same form as in 
\cite{bhmr91,jow87,rug90}. The derivations
based on the semiclassical radiation theory in \cite{dk75} are very arduous
but unavoidable for the ST part of the picture. However, one is tempted to    
try to obtain the pure spin diffusion part via the traditional route based
on Gaussian white noise in analogy to the description of orbital motion.
This would lead to a better appreciation of the results in \cite{dk75} and
to more  insights. We have succeeded in this approach and 
proceed by developing  our arguments within a purely 
classical framework in which $\vec S$ is treated as a classical spin vector.
We therefore postpone further discussion of the semiclassical calculation 
until later.

\section{Spin--orbit transport without radiation}
In the absence of radiation and other non--Hamiltonian effects 
and with the orbital Hamiltonian ${h}_{\rm orb}$,  the orbital 
phase space density $W_{\rm orb}$ evolves according to an equation of the
Liouville type:
\begin{eqnarray}
\frac{\partial W_{\rm orb}}{\partial s}~ 
              =~ \lbrace{{h}}_{\rm orb} ,W_{\rm orb} \rbrace_{\vec u}
\end{eqnarray}
where ${\lbrace{\, , }\rbrace}_{\vec u}$ is the Poisson bracket involving
derivatives w.r.t. the components of $\vec u$.
We normalise the density to unity: $ \int d^6 u ~ W_{\rm orb}(\vec u; s) = 1$.

Since the T--BMT equation is linear in spin, the local polarisation
$\vec P_{\rm loc}(\vec u; s)$, 
which is proportional to an average over spin vectors in an infinitesimal
packet of phase space at $(\vec u; s)$, obeys the T--BMT equation 
along any orbit $\vec u(s)$. 
If $\vec P_{\rm loc}(\vec u; s)$ is a smooth function of $(\vec u; s)$ 
we can rewrite this as
\begin{equation}
  \frac{\partial {\vec{P}}_{\rm loc}}{\partial s}~ =~
      \lbrace { h}_{\rm orb},  {\vec{P}}_{\rm loc}  \rbrace_{\vec u}
        +   \vec{\Omega}  \wedge   {\vec{P}}_{\rm loc} \; .
\end{equation}

\section{Spin--orbit transport with radiation}
To include radiation we model the photon emission as a Gaussian
white noise process overlaid onto smooth radiation damping. Then  eq.~(2) 
is replaced by a FP equation:
\begin{eqnarray}
\frac{\partial W_{\rm orb}} {\partial s}   =
       {\cal L}_{_{\rm FP,orb}} \; W_{\rm orb} \; ,
\end{eqnarray}
where the orbital FP operator can be decomposed into the form:
\begin{eqnarray}
{\cal L}_{_{\rm FP,orb}} = {\cal L}_{\rm ham} + {\cal L}_{0} + {\cal L}_{1} + 
{\cal L}_{2} \, 
\end{eqnarray}
and where ${\cal L}_{\rm ham}$ would result in eq.~(2) and 
${\cal L}_{0},~ {\cal L}_{1},~{\cal L}_{2}$ are terms due to damping and noise
containing zeroth, first and second order derivatives w.r.t. the components of
$\vec u$ respectively.
The detailed forms for the $\cal L$'s can be found in 
\cite{bhmr91,jow87,rug90}
but are not important for the argument that follows. After a few damping times
$W_{\rm orb}$ approaches an equilibrium form.

But how can we write the analogue of eq.~(4) for polarisation?  After all, to
obtain an equation of the FP type we need a density and 
polarisation is not a density. But we do have the 
spin angular momentum density.
In particular we have the {\it spin angular momentum density per particle}, 
$\vec {\cal S}$, and its close relative the
{\it polarisation density} $\vec{\cal P} = 2/\hbar~ \vec {\cal S}$.
This latter can be written as
\begin{eqnarray}
{\vec {\cal P}}(\vec u; s) =
 {{\vec P}_{\rm loc}}(\vec u; s)~W_{\rm orb} (\vec u; s) \; . 
\end{eqnarray}
By combining eqs.~(2) and (3) we then obtain
\begin{eqnarray}
 \frac{ {\partial \vec{\cal P}}}{\partial s} =
      \lbrace {h}_{\rm orb},  \vec{\cal P}  \rbrace_{\vec u}
        +   \vec{\Omega}  \wedge   \vec{\cal P} \; .
\end{eqnarray}
This equation for the polarisation density has the same form as eq.~(2)
for the phase space density except for the precession term and since eq.~(2)
is just the radiationless version of eq.~(4) we can now guess how
the extension of eq.~(7) to include radiation will look.

To come further we parametrise the spin components in terms of the 
canonical variables
$J~~ {\rm and~~} \psi $ defined by the relations
{}~$ S_1 = \cos(\psi)\sqrt{{{\hbar}^2}/{4} -J^2}$,~ 
$ S_2 = \sin(\psi)\sqrt{{{\hbar}^2}/{4} -J^2}$ ~and~ 
$ S_3 = J$~
and having the Poisson bracket 
$ \lbrace \psi, J \rbrace_{\psi, J} = 1 $ \cite{yok86,bhrnotes,bhr1}.
 These lead to the standard 
Poisson brackets for angular momentum: 
$ \lbrace { S}_j , { S}_k\rbrace =
    \sum_{m=1}^3 \varepsilon_{jkm}~ { S}_m$.
The spin variables commute with the orbital variables. 

In terms of the 
combined spin--orbit  Hamiltonian $h = h_{orb} + {\vec S}\cdot{\vec {\Omega}}$
the T--BMT equation can now be written as 
{}~~$d \vec {S}/{ds}= \lbrace {\vec S} , h  \rbrace_{{\vec u}, \psi, J}$
and the equations of radiationless orbital motion 
{}~~$d \vec {u}/{ds}= \lbrace {\vec u} , h \rbrace_{{\vec u}, \psi, J}$,
are the usual equations of orbital motion except for additional 
terms accounting for Stern--Gerlach (SG) forces.

We now need the joint  spin--orbit density 
$W(\vec u, \vec S; s)$. 
This contains a factor $\delta(\hbar/2-|{\vec S}|)$ to account for the fact 
that we wish to describe processes for which $|\vec S| = \hbar/2$
and we normalise $W$ to unity: 
$ \int d^6u~ d^3S ~ W(\vec u,{\vec S}; s) = 1$.
Moreover $\int d^3S ~ W(\vec u,{\vec S}; s) =  W_{\rm orb}(\vec u; s)$.

Equation~(6) for the polarisation
density can then be written as
\begin{eqnarray}
{\vec {\cal P}}(\vec u; s) = 
           \int d^3S ~ \frac{{\vec S}}{|{\vec S}|} ~W(\vec u, \vec S; s) \; . 
\end{eqnarray}
The polarisation of the whole beam as measured by a polarimeter at azimuth
$s$ is 
${\int} d^6 u~ {\vec {\cal P}}({\vec u}; s) $.

Since here, spin is a spectator, being only indirectly affected by the
 radiation through the orbital motion,
the FP equation for the combined orbit and spin density is
\footnote{The critical energy for synchrotron radiation is usually tens of 
KeV but the SG
energy is many orders of magnitude smaller \cite{mont98I}. Therefore the
influence of spin motion on the orbital motion can be neglected.
To include the SG forces one replaces
the Poisson bracket term in eq.~(9) with
{}~~$\lbrace \vec{\Omega}\cdot\vec S \; , W \rbrace _{{\vec u}, \psi, J}$.}
\begin{eqnarray}
&& \frac{\partial W}{\partial s} = {\cal L}_{{}_{\rm FP,orb}} W 
-  (\vec {\Omega} \wedge \vec {S}) \cdot (\vec\nabla_{_{\vec S}} W) = 
{\cal L}_{{}_{\rm FP,orb}} W
+  \vec{\Omega}\cdot \lbrace \vec S \; , W \rbrace_{ \psi, J}  
\end{eqnarray}
where $\vec\nabla_{_{\vec S}} W$ is the gradient of $W$ w.r.t. the three
components of spin. The spin part of the corresponding Langevin equation 
has no noise terms.

Using  eq.~(9) we can write
\begin{eqnarray}
 \int d^3S ~\frac{{\vec S}}{|{\vec S}|}~ \frac{\partial W}{\partial s}
 = \int d^3S ~\frac{{\vec S}}{|{\vec S}|} \biggl({\cal L}_{{}_{\rm FP,orb}} W 
+ \vec{\Omega}\cdot\lbrace \vec S \; , W \rbrace_{\psi, J}\biggr) \;
\end{eqnarray}
and then by eq.~(8) we obtain
\begin{eqnarray}
 \frac{\partial \vec{\cal P}}{\partial s}~ =~
             {\cal L}_{_{\rm FP,orb}}\vec{\cal P} 
        +   \vec{\Omega}  \wedge   \vec{\cal P}  \; .
\end{eqnarray}

This is the extension of eq.~(7) to include radiation that we have been
seeking and we see that it is an obvious generalisation of eq.~(7).
If we switch off the radiation, we of course obtain eq.~(7) but by  
introducing $W ({\vec u}, {\vec S}; s)$ we avoid the heuristic 
derivation of eq.~(3). We call eqs.~(7) and (11) ``Bloch'' equations
following the usage for equations of this general form 
in the nuclear magnetic resonance  literature. Concrete examples 
of eqs.~(4)--(11) for simple 
exactly solvable models can be found in \cite{kh97}.

\section{Discussion and conclusion}
The derivation of the Bloch equation for $\vec{\cal P}$ given here is 
independent of the source of noise and damping. In fact as soon as we have the 
${\cal L}_{_{\rm FP,orb}}$ for a process we can write down the 
corresponding Bloch equation for $\vec{\cal P}$. 
Furthermore,  providing that spin is a spectator, this approach can be 
applied to more general 
diffusion problems where the operator ${\cal L}_{_{\rm FP,orb}}$ is  
replaced by the appropriate form. 
For example, a Bloch equation can be written to describe the effect of 
intrabeam scattering without spin flip or the 
scattering of protons without spin flip off gas atoms and molecules. 
Note that  the Bloch 
equation is valid far from spin--orbit equilibrium and that it is linear
in $\vec{\cal P}$. Moreover, it is universal in the sense that it does not 
explicitly contain the
orbital density $W_{\rm orb}$. Surely this is the best place to begin
discussions on spin diffusion.
In the case of noise and damping due to 
synchrotron radiation and if the spin--orbit coupling term in eq.~(11) were
to vanish ($\vec \Omega = \vec 0$), the three components of $\vec{\cal P}$
 would
each reach equilibrium forms proportional to the equilibrium form for 
$W_{\rm orb}$. However $\vec \Omega $ does not vanish but instead   
mixes the components. This is the route, in this picture, by which 
$\vec \Omega $ causes depolarisation. 

The corresponding evolution equation for the local polarisation
 ${\vec P}_{\rm loc}$
can be found by substituting eq.~(6) into eq.~(11) and using eq.~(4)
but the resulting equation is 
complicated owing to the second derivative in ${\cal L}_{2}$ and it is 
{\it not}
universal since it contains $W_{\rm orb}$.
So to extract ${\vec P}_{\rm loc}$ one should first solve eqs.~(4) and (11)
separately and then use eq.~(6).

The semiclassical calculation \cite{dk75} involves writing the density 
operator in two component spin space as 
$\rho = \frac{1}{2}( \rho_{\rm orb} + \vec {\sigma} \cdot \vec{\xi})$ where 
$\vec {\sigma}$
is the spin operator, $\rho_{\rm orb}$ is the density operator of the
orbital motion  and where the operator  $\vec{\xi}$, which encodes information
about the polarisation, is  
equivalent to  $\vec{\cal P}$.
In the quantum mechanical picture, all expectation values 
involving spin can depend only on $\vec{\xi}$ and different mixed spin states
leading to the same $\vec{\xi}$ are equivalent.   
Correspondingly,
the definition of ${\vec{\cal P}}(\vec u; s)$ (eqs.~(6) and~(8))
involves integration over the spin distribution at $(\vec u; s)$ so that
in principle different spin distributions at 
$(\vec u; s)$ can lead to the same ${\vec{\cal P}}(\vec u; s)$.
Thus ${\vec{\cal P}}$ is  not only  an economical representation 
of the spin motion by virtue of its being an average over spin degrees of 
freedom, but
even as a classical entity it also embodies the effective
indistinguishability of equivalent spin distributions.      

At zeroth order in $\hbar$ and in the absence
of radiation the Weyl transform of $\vec \xi$ fulfills eq.~(7) as expected
\cite{mont98I, khthesis}. At higher order in $\hbar$~ SG effects appear.  
The corresponding calculation in the presence of radiation \cite{dk75} 
delivers terms equivalent to those on the r.h.s. of eq.~(11),
which are due to pure spin diffusion, together with terms due to the ST effect
which are, not surprisingly, of the Baier--Katkov--Strakhovenko form 
\cite{bks70,mont84}. There are also the cross terms. 
So starting with eq.~(11) one
could, on the basis of physical intuition, add in the ST terms by hand.
But the cross terms can be very important \cite{mont98I} and they would be 
missed. So to obtain a complete description of spin motion a full quantum 
mechanical, or at least semiclassical, treatment of combined spin and 
orbital motion is unavoidable.  
Our work is a classical reconstruction of the pure noise and damping part of
eq.~(2) in \cite{dk75}. Since the evolution equation for the orbital phase 
space density in \cite{dk75} is the usual FP equation, one sees that the 
calculation in \cite{dk75} provides a physical justification for using
Gaussian white noise models for orbital motion. The use of eq.~(11)
and of eq.~(2) in \cite{dk75} will be described in a future paper.

It should now be clear that the polarisation density is the most natural    
polarisation--like quantity to use in FP--like descriptions of 
spin motion in accelerator physics. In fact in retrospect its (three 
component) equation of motion (eq.~(11)) is an intuitively 
obvious generalisation of the (one component) equation for the
particle density (eq.~(4)) with an extra term to describe the T--BMT 
precession of the polarisation density. 
Moreover, since the spin degrees of freedom
have been integrated out, the problems of dealing with FP equations
containing (spin) variables describing motion on the sphere and of enforcing 
periodicity conditions for the spin distribution, are bypassed.
Perhaps some problems in condensed matter physics involving spin diffusion
due to fluctuating magnetic fields could be conveniently handled by simulating
the field fluctuations in terms of particle motion in an artificial ``phase
space'' and then working with the accompanying artificial polarisation density.

\section*{Acknowledgments}
We thank M. Berglund for careful checking of the manuscript.

\end{document}